\documentclass[10pt,letterpaper,authoryear]{elsarticle}
\usepackage{graphicx}
\usepackage{rotating}
\usepackage{nameref}
\usepackage{bm}
\usepackage{amssymb}
\usepackage{hyperref}
\usepackage{cite}
\usepackage{amsmath}
\usepackage{natbib}
\usepackage{dcolumn,multicol}

\begin{document}
%
%
%
\vspace*{0.35in}

\begin{flushleft}
{\Large
\textbf\newline{Changes in sinking of plankton--like particle: comparison between observations and numerical model}
}
\newline
\\
E.Y.Shchekinova\textsuperscript{1,2},
Christina Geb{\"u}hr\textsuperscript{1,3}, 
Maarten Boersma\textsuperscript{1}, 
Karen H. Wiltshire\textsuperscript{1}
\\
\bigskip
\bf{1} Alfred--Wegener--Institut Helmholtz--Zentrum f{\"u}r Polar-- und Meeresforschung, Biologische Anstalt Helgoland, Kurpromenade 201, D-27498 Helgoland, Germany
\\
\bf{2} International Institute for Advanced System Analysis, Schlossplatz 1, A-2361 Laxenburg, Austria
\\
\bf{3} Institute of Geosciences, Goethe--University Frankfurt, 60438 Frankfurt am Main, Germany\\
\bigskip
* shchekinova.elena@gmail.com

\end{flushleft}

\section*{Abstract}
Sinking and resuspension are the most important processes determining the distribution of bentho--pelagic diatoms in the water column and influencing their variability. From the bentho--pelagic time series $1991-2008$ we analyzed  weekly concentrations of the diatom \textit{Paralia sulcata} in the German Bight. The analysis was carried out for lower wind ranges ($<10$ m s$^{-1}$) at two different seasons characterized by distinct sea surface conditions. Our results showed that the relationship between wind and diatom concentrations was much influenced by seasonal conditions: the decrease of abundances at higher winds was more pronounced in summer. Additional sampling procedure of species at two different water depths (bottom and surface layers) in the years $2007-2009$ showed that in summer the surface abundance of species decreased substantially with increasing wind. In contrast, in winter the decrease of the surface concentration of species was less pronounced. To explain observed seasonal variations  numerical simulations of plankton--like particles were performed using a two--dimensional turbulence model. Numerical simulations were carried using parameters corresponding to averaged winter and summer temperature and salinity at the study site. We showed that the reduction of cell counts was possibly linked to the changes in sinking of diatoms due to the altered buoyancy of algal cells. Our numerical results identified reduction of sinking speed of particles for the parameters corresponding to winter conditions.  
%
\section*{Introduction}\label{sec.1}
Together with tides, wind is one of the most important environmental drivers of turbulence and mixing processes in coastal seas \citep{Young1999, Dejonge1995, Demers1987, Song2011}. Especially in shallow waters the wind energy causes vertical mixing, often through to the sea floor \citep{Weisse2009}. As a consequence, life on the shelf is highly dependent on wind induced mixing processes \citep{Demers1987}. In particular, stormy conditions and wind mixing are responsible for the stirring and uplifting of sediments as well as organisms such as benthic diatoms into the water column \citep{Therriault1978, Levasseur1983, Nichols1985, McQuoid2003a, McQuoid2003b, Iriarte2004}. Moreover, in the shallow areas wind obviously also affects water clarity and underwater light climate, and as such the short--term variability of wind speeds can impact the timing of algal blooms via variations in turbidity, tidal fronts and surface mixed layer depth as exemplified for the spring bloom of diatoms in the southern North Sea by Tian et al.(2011). 

Many algal species occur both in the water column as well as on the sediment of shallow seas, because they sediment out of the water column during quiet periods, and are resuspended during periods of turbulence. Some species, such as, for example, the diatom \textit{Paralia sulcata} (Ehrenberg) Cleve 1873, form microphytobenthal mats, and leave these mats in periods of higher turbulence \citep{Gebuehr2009}. 

The relation between wind and the concentrations of bentho--pelagic algae in the water column was addressed previously for estuaries \citep{Dejonge1995, Demers1987}. In the study of microphytobenthos in the Ems estuary \citep{Dejonge1995} the fraction of the benthic organisms moving from the top of sediment into the water column was linearly dependent on the effective wind speeds. In contrast, the analysis by Demers et al. (1987) on resuspension of organic particles in the littoral zone of the St. Lawrence estuary showed that a critical wind velocity was needed to cause a significant increase of particulate organic matter in the water column, and hence a non--linear response to wind was observed.

Previous observational studies \citep{Roelofs1984, Hobson1997} showed positive correlations between increased wind and number of the benthic organisms that enter the water column from sediments. Moreover, the onset of high wind conditions in autumn is linked to the re--appearance of bentho--pelagic diatoms in the water column \citep{Abrantes1988, McQuoid1998, Gebuehr2009}, but an abrupt decrease of surface concentrations of diatoms such as \textit{P. sulcata} sometimes was observed during storms \citep{Gebuehr2011}. Thus, the dependence of the diatom concentrations on wind conditions remains not well understood. 

In the studies of sedimentations under different temperature conditions \citep{Kroegel1998} it was shown that sedimentation processes rely on the water kinematic viscosity adjustment and higher settling velocity of suspended particulate matter was observed in summer. Another study on sinking rates \citep{Bienfang1982} showed that diatom cell buoyancy changes according to variations in sea surface temperature and salinity. This also implies that depending on the seasonal water conditions, wind may have a different impact on the sedimentation behavior of bentho--pelagic algae, which potentially could explain the somewhat varying results obtained from the literature \citep{Roelofs1984, Hobson1997,Gebuehr2011}. The sinking behavior of a diatom cell \citep{Laennergren1979} could depend on variable sea surface conditions, specifically, it could be affected by seasonal changes of cell buoyancy and fluid density. This could occur due to seasonal variations in salinity and temperature.

Here, we investigated the influence of seasonal sea surface conditions, wind magnitude and densities of the bentho--pelagic diatom \textit{P. sulcata} (Ehrenberg) Cleve in the water column at Helgoland in the North Sea. Our study is focused on lower weekly wind ranges ($<10$ m s$^{-1}$) and observations during stormy weather conditions were not considered here. We used simultaneous long--term wind data and diatom cell counts from the same observational site. To evaluate the effect of seasonal variations of sea surface salinity and temperature on the sinking behavior of diatoms under turbulent conditions we performed a set of numerical simulations. We used a two--dimensional turbulence model with the mean temperature and salinity values from the observational data at Helgoland. The results were compared with the measurements of cell counts of \textit{P. sulcata} obtained for the whole water column and for two distinct sea depth levels at different sea conditions.

 \section*{Materials and Methods}
 \subsection*{Description of study site}
For the estimates of plankton densities we used the Helgoland Roads (HR) algal data \citep{Radach1986, Franke2004, Wiltshire2004b}. Helgoland is located in the German Bight about $65$ km off the German coast (Fig. \ref{fig1}) in the confluence of the estuaries of the rivers Weser and Elbe. Thus, the area is influenced by the circulation of the low saline coastal waters and the open waters of the southeastern North Sea \citep{Bauerfeind1990, Ionita2008}. At HR the water is well--mixed due to tidal activity \citep{Hickel1998} and shallow depths (between $3$ and $10$ m) \citep{Dekluijver1991, Wiltshire2010}. The hydrographic situation at Helgoland is highly variable due to close proximity of the coastal waters and also due to variable wind conditions \citep{Dippner1993, Dippner1998, Siegismund2001, Tian2011}.  
 
\subsection*{Description of the \textit{P. sulcata} dataset}
 
\textit{P. sulcata} is representative (in terms of abundances and regularity in the data set) of the benthic--pelagic diatom community at HR \citep{Wiltshire2004b, Gebuehr2009}. This species is a centric diatom that is regularly found in the sediment and in the water column in the North Sea \citep{Drebes1974, Hoppenrath2009,Kraberg2010}. In general, it is more common in the benthos than in the pelagic zone \citep{Sancetta1989}. In other areas of the world \textit{P. sulcata} is usually found in zones with upwelling \citep{Abrantes1988}. 

For \textit{P. sulcata} continuous yearly records are available since $1962$. In the study on \textit{P. sulcata} response to various environmental factors by Geb{\"u}hr et al. (2009) it was shown that the species is better adjusted to colder temperatures and lower light availability. Starting from $1998$ the species has been found on a regular basis throughout the summer (Fig. \ref{fig2}) \citep{Wiltshire2004a,Gebuehr2009} . 

In the long--term sampling, surface water samples were taken on every working day (Monday--Friday) except for holidays. The details of phytoplankton water sampling procedure and identification were described in Wiltshire and D{\"u}rselen (2004) and in Geb{\"u}hr et al. (2009). Microalgae were identified to the species level. We used weekly means of the abundance data to avoid variability and to fill the gaps of missing data. Due to seasonal variability of weather and different environmental factors that influence diatom growth as well as the variability of HR daily diatom counts, we sorted data into winter (Dec--Feb) and summer (Jun--Aug) sets. Indeed, winter conditions at Helgoland are characterized by higher nutrient concentrations. In contrast, in summer, warmer water temperatures and higher light levels as well as limiting nutrients characterize the environment. 

The abundances of  \textit{P. sulcata} typically correlated with periods of water turbulence \citep{Kraberg2010}. The species was often found in the plankton after storms due to strong vertical mixing of the water column \citep{McQuoid1998}.

The weekly means and standard deviations of the species abundances were estimated and sorted according to the weekly wind speed evaluated in the week preceding to the algae measurements. The abundance data were grouped into two wind ranges ($0\le W_{mean}< 5$ m s$^{-1}$, $5\le W_{mean}<10$ m s$^{-1}$). Thus, we accounted for the changes in algae concentrations that could be affected by wind conditions in the previous week. Data were grouped separately for summer and winter observations. 

\begin{figure}
       \scalebox{0.3}{\includegraphics{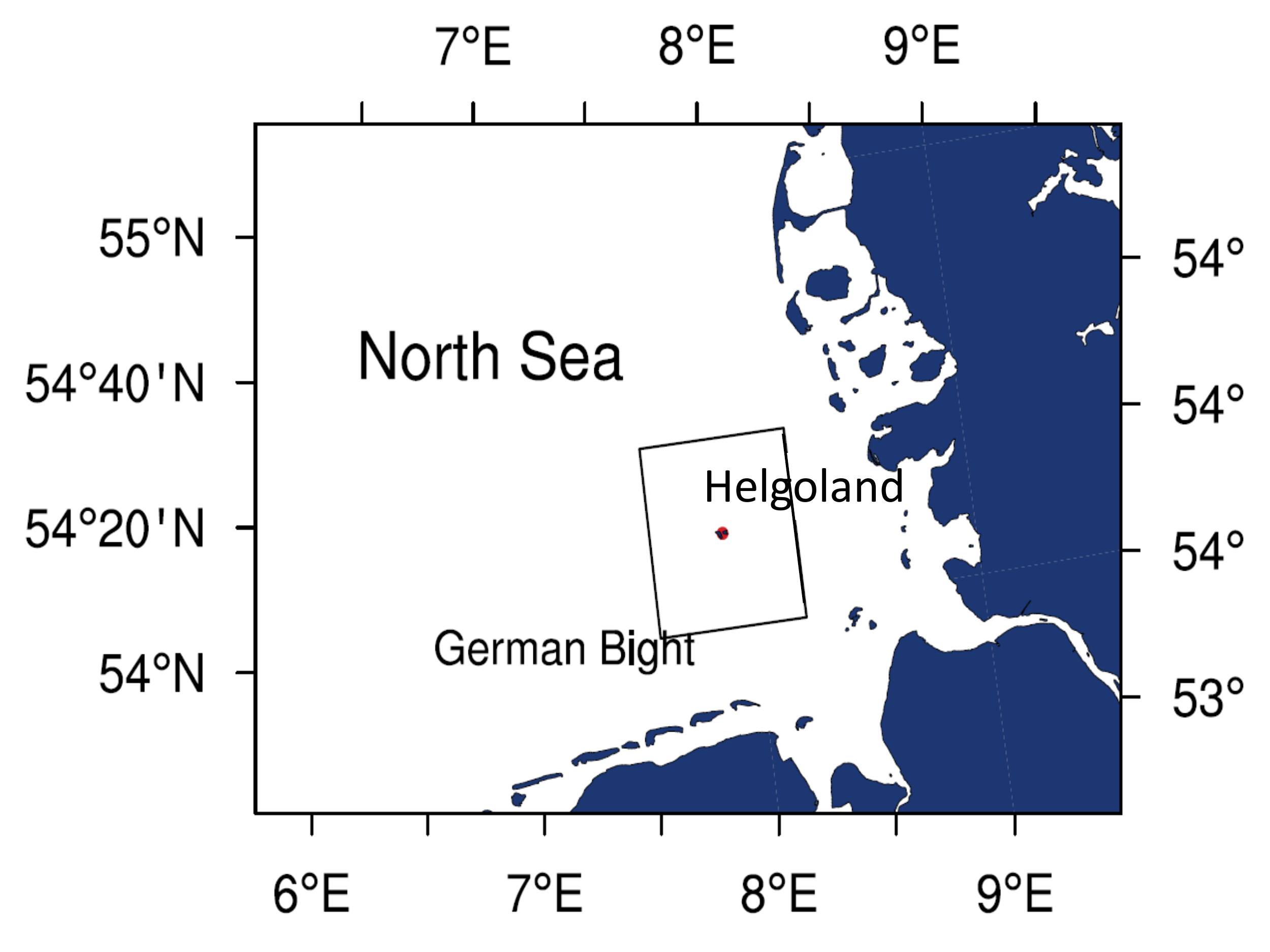}}
     \caption{Location of the study site in the German Bight, North Sea and Helgoland Roads (inside box) located at ($54^\circ 11'$ N, $7^\circ 54'$ E ).}\label{fig1}
   \end{figure}

Additionally to long--term sampling a weekly water sampling (in total $104$ samples) was conducted at the same sampling site at HR (see Fig. \ref{fig1}) to investigate the occurrence of \textit{P. sulcata} at $1$ m above the sediment and at surface level during the period $16.10.2007-29.10.2009$ \citep{Gebuehr2011}. The samples were taken simultaneously from the bottom and surface depth levels (the depth at the sample site is within the range $6-10$ m ). The bottom and surface densities of \textit{P. sulcata} were sorted according to seasons (winter and summer) and wind ranges. 
\subsection*{Description of wind data}
We used $10$ m atmospheric wind measurements of the German Meteorological Service for Helgoland. The wind measurements from Helgoland site were used in the past as a proxy for studies of regional ecosystem processes \citep{Tian2011} and energy productions of offshore wind turbines \citep{Lindenberg2012}. The mean daily winds were obtained as arithmetic means over at least $21$ hours of hourly wind measurements. To evaluate the wind activity in recent decades ($1991-2008$) the weekly means of the daily wind measurements were analyzed. This allows avoiding spurious fluctuations on a daily scale. Here we concentrated on weekly winds below $10$ m s$^{-1}$. 
Unfortunately, we were forced to use a shorter time series than is available for phytoplankton because the wind station at Helgoland was moved in $1989$ \citep{Lindenberg2012}, and a fairly large discrepancy exists between data before and after this move. Although the exact measuring procedure was altered several times in the period since $1989$ these changes did not affect wind data consistency \citep{Lindenberg2012}.
\section*{Model} 
To describe diatoms vertical mixing we used a two--dimensional turbulence model for finite--sized, non--neutrally buoyant particles \citep{Fung2003, Fung2008}. For a heavy and spherical particles with a radius smaller than the Kolmogorov length scale only the drag and buoyancy forces are important. Based on the above assumptions the equation of motion is written as follows:
\begin{equation}
\frac{d\mathbf{v}}{dt}=\frac{1}{\tau_p}[\mathbf{u}(\mathbf{x}_p,t)-\mathbf{v}(t)]+\mathbf{g},\label{equationMotion}
\end{equation}
where $\mathbf{v}(t)=d\mathbf{x}_p/dt$ is two--dimensional turbulence velocity field, $\tau_p$ is relaxation time defined from the expression:
\begin{equation}
\tau_p=\frac{2\rho_p a^2}{9\rho_f \nu}, 
\end{equation}  
where $\rho_p$ and $\rho_f$ are algal cell and water densities respectively, $a$ is radius of the diatom and $\nu$ is the water kinematic viscosity. This velocity field is chosen as in Fung and Vassilicos, 2003 (see Appendix). The model of turbulent flow was used previously to investigate settling of heavy particles and the mechanism of prolonged suspension of particles in a turbulent fluid \citep{Pasquero2003}.  

The density parameters of sea water depending on salinity and temperature were evaluated from the equation of state for sea water \citep{Millero1981}. We used mean winter and summer salinity and temperature values obtained from the long--term times series at HR: $(T_{win}= 5.27^\circ \text{C}, S_{win}= 32.42$ PSU and $T_{sum} = 16.32^\circ \text{C}, S_{sum}=32.19$ PSU). The values for the summer and winter kinematic viscosities are $\nu_{sum}=1.1092\times 10^{-6}$ m$^2$ s$^{-1}$ and $\nu_{win}=1.519\times 10^{-6}$ m$^2$ s$^{-1}$ correspondingly. The mean radius of the diatom frustule is $10$ $\mu$m and the average density of algal cell $\rho_p=1300$ kg m$^{-3}$. We considered particles initialized in a square domain with a homogeneous, isotropic and statistically stationary random velocity field with periodic boundary conditions. The total number of $200$ particles was simulated in a 2D box with open boundaries for  time period $t=20\pi/\min(\omega_n)$, where $\min(\omega_n)$ is the minimal wave frequency. Two sets of numerical tests were performed using summer and winter parameters. 
\section{Results} 

   \begin{figure}
    \scalebox{0.4}{\includegraphics{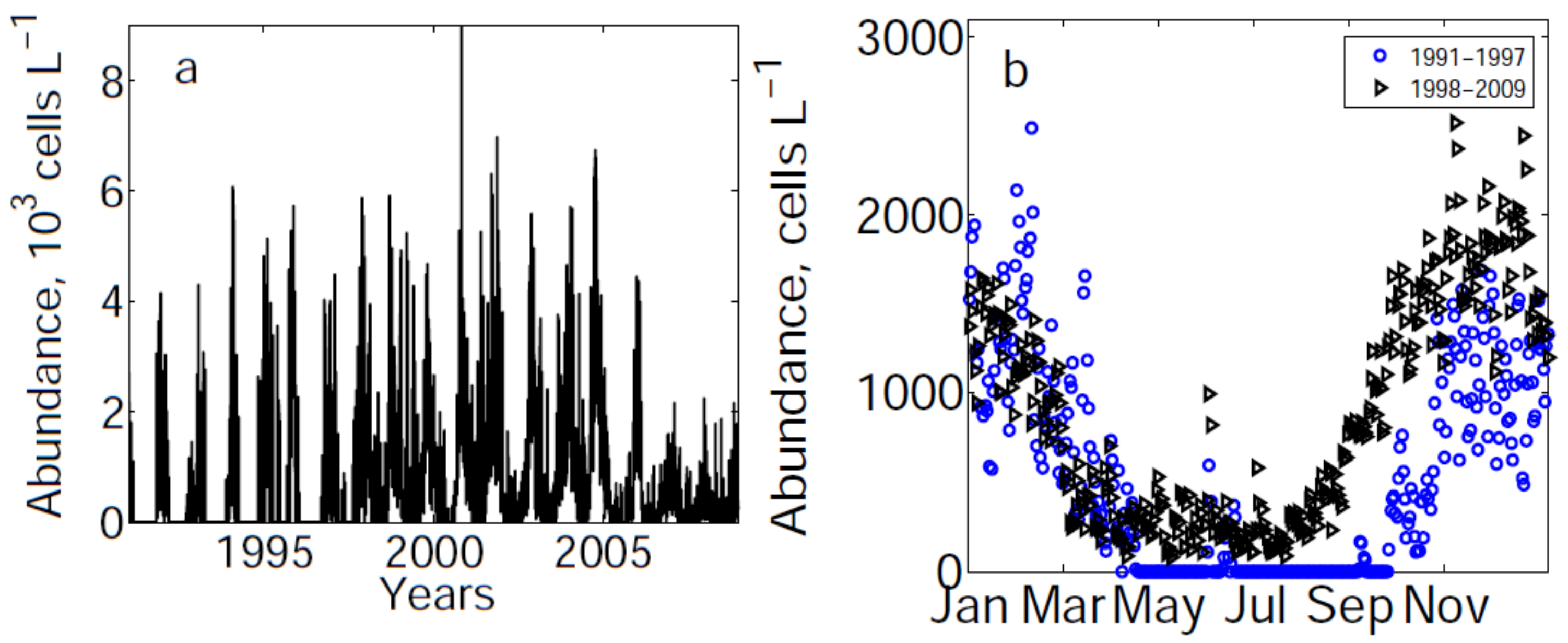}}
      \caption{(\textbf{a}) Abundance for \textit{P. sulcata} from the HR data set for period $1991-2008$. (\textbf{b}) Mean abundance of \textit{P. sulcata} for the years $1991-1997$ (circles) and $1998-2009$ (triangles). Data were calculated based on the time series of weekly means of diatom abundances. }\label{fig2}
    \end{figure}
      \begin{figure}
        \scalebox{0.2}{\includegraphics{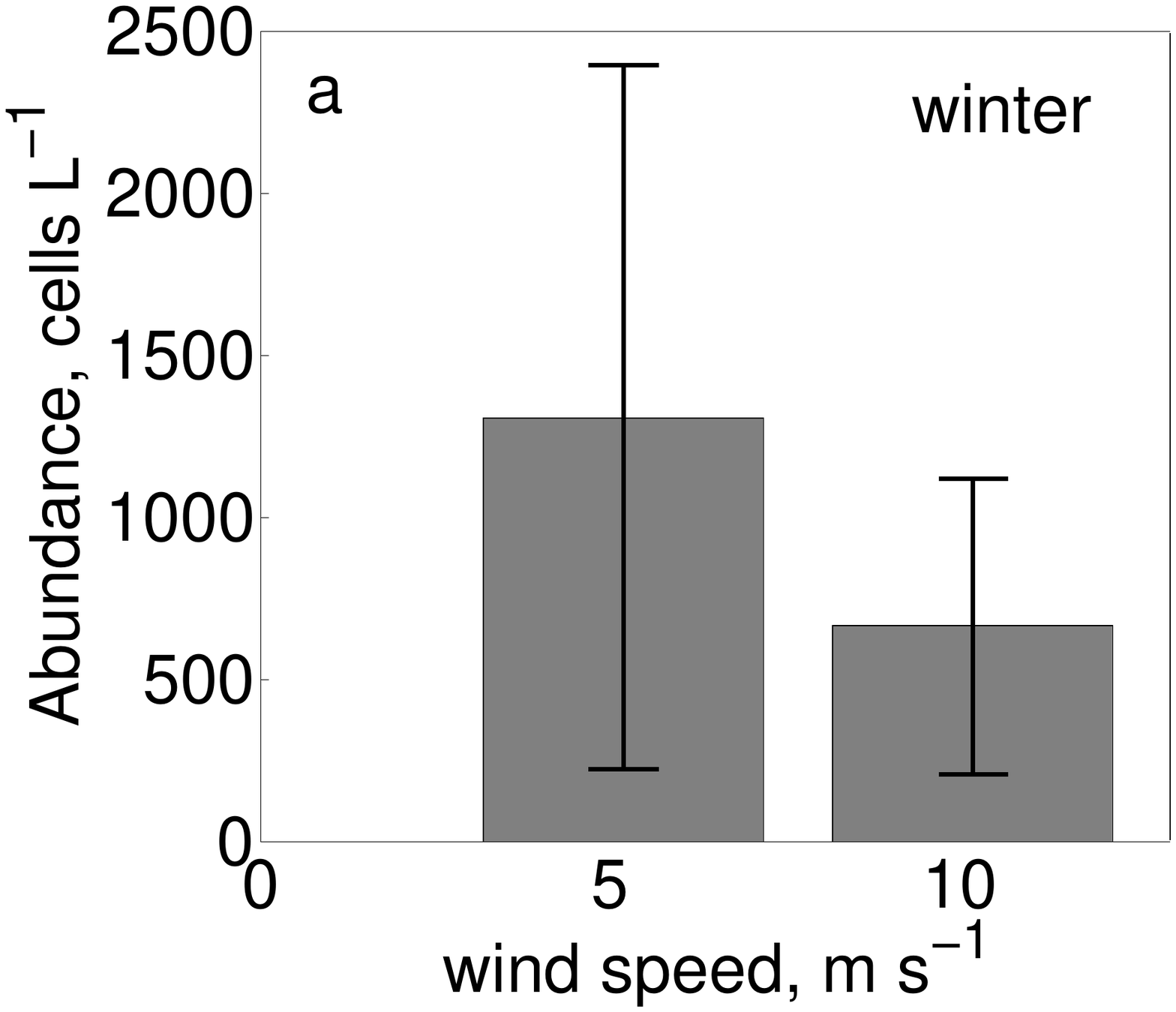}}
        \scalebox{0.2}{\includegraphics{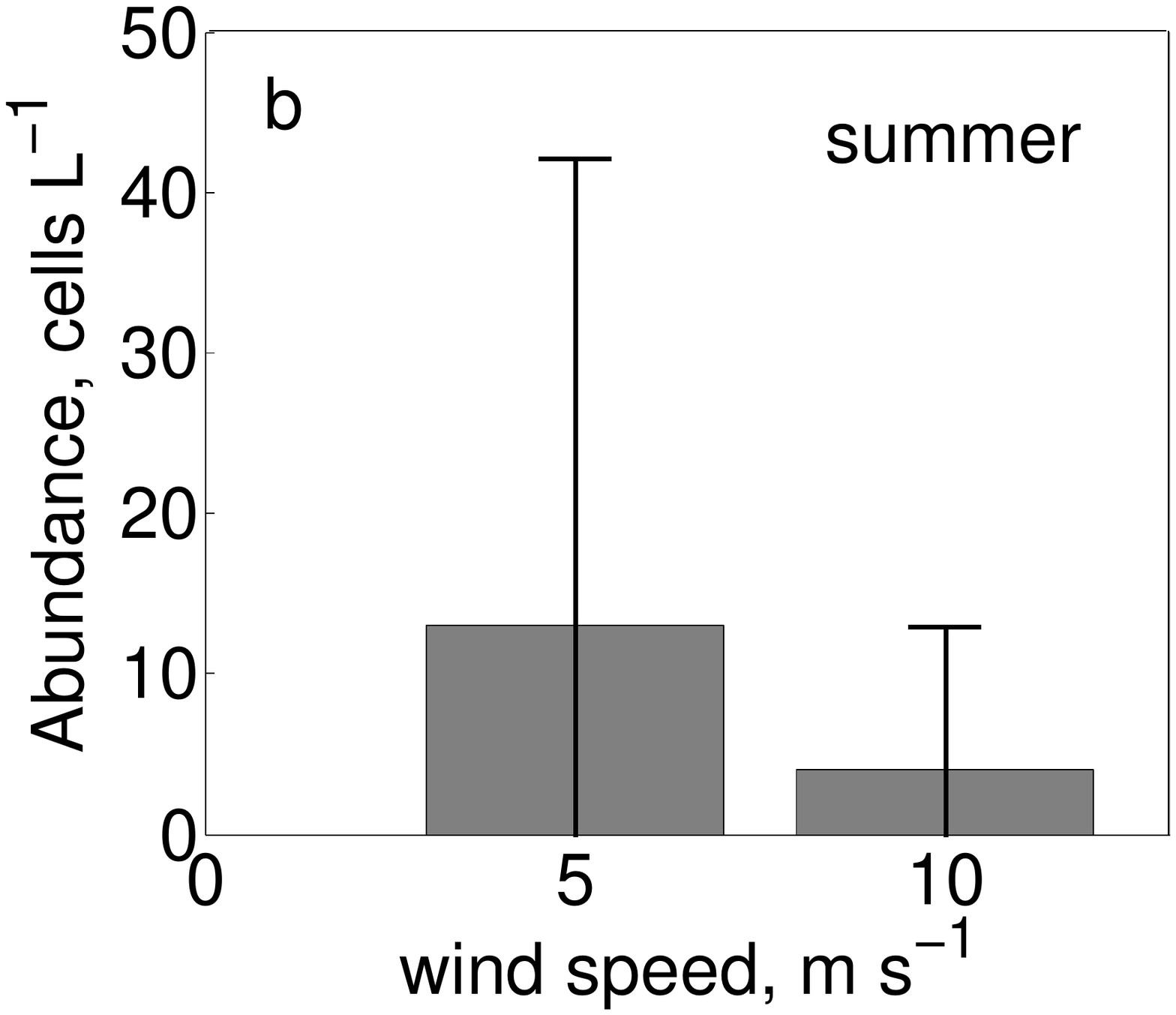}}
          \caption{Mean and standard deviations of abundances [cells L$^{-1}$] of \textit{P. sulcata} for winter (\textbf{a}) and summer (\textbf{b}) for two wind ranges: $(1)$ $[0, 5]$ m s$^{-1}$, $(2)$ $[5, 10]$ m s$^{-1}$. Data were regrouped using time series of weekly means of wind and diatom abundances for period $1991-2008$. The abundances were sorted according to wind ranges observed in the week preceding the sampling of algae.}\label{fig3}
        \end{figure}
     \begin{figure}
        \scalebox{0.2}{\includegraphics{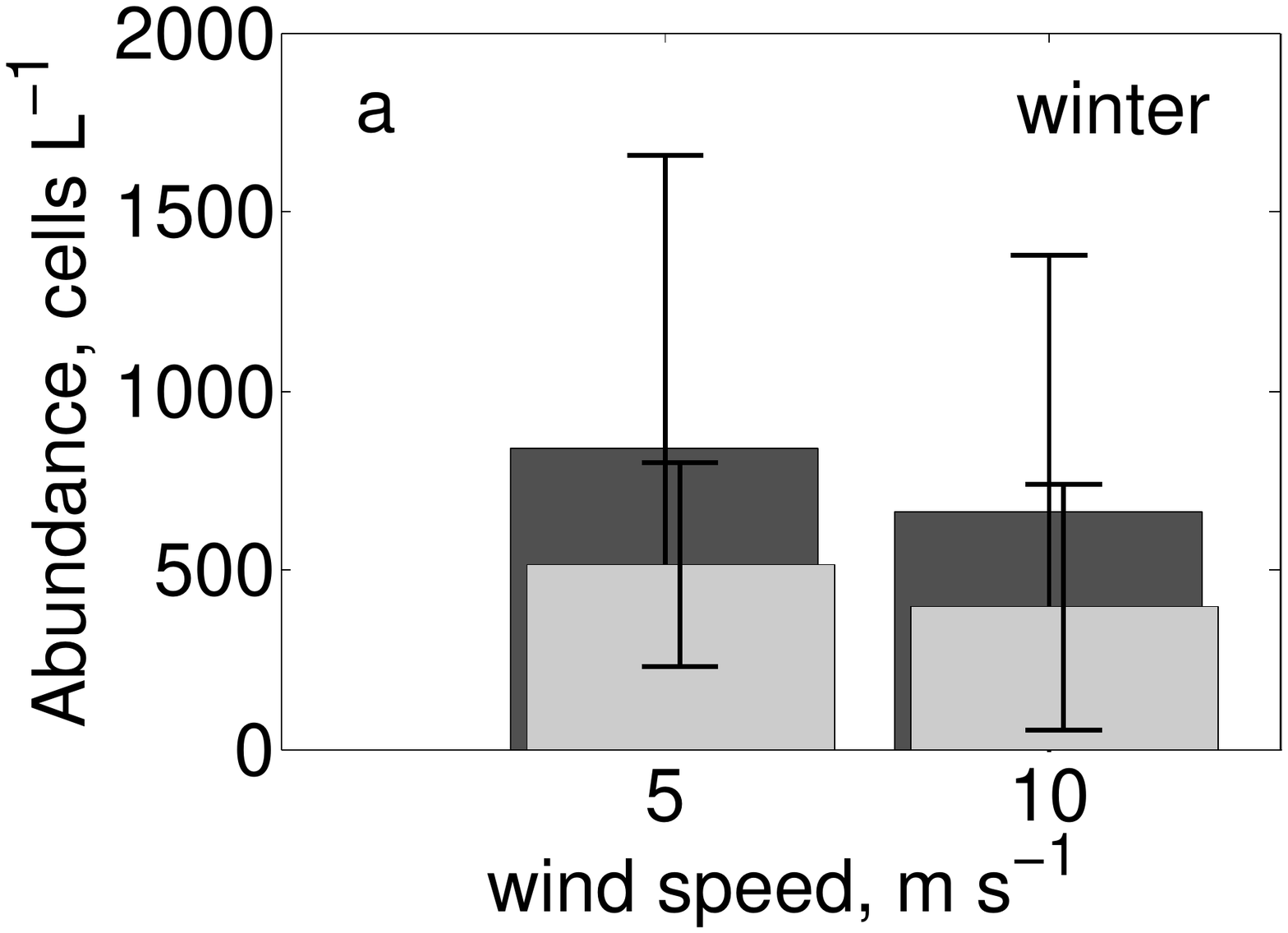}}
        \scalebox{0.2}{\includegraphics{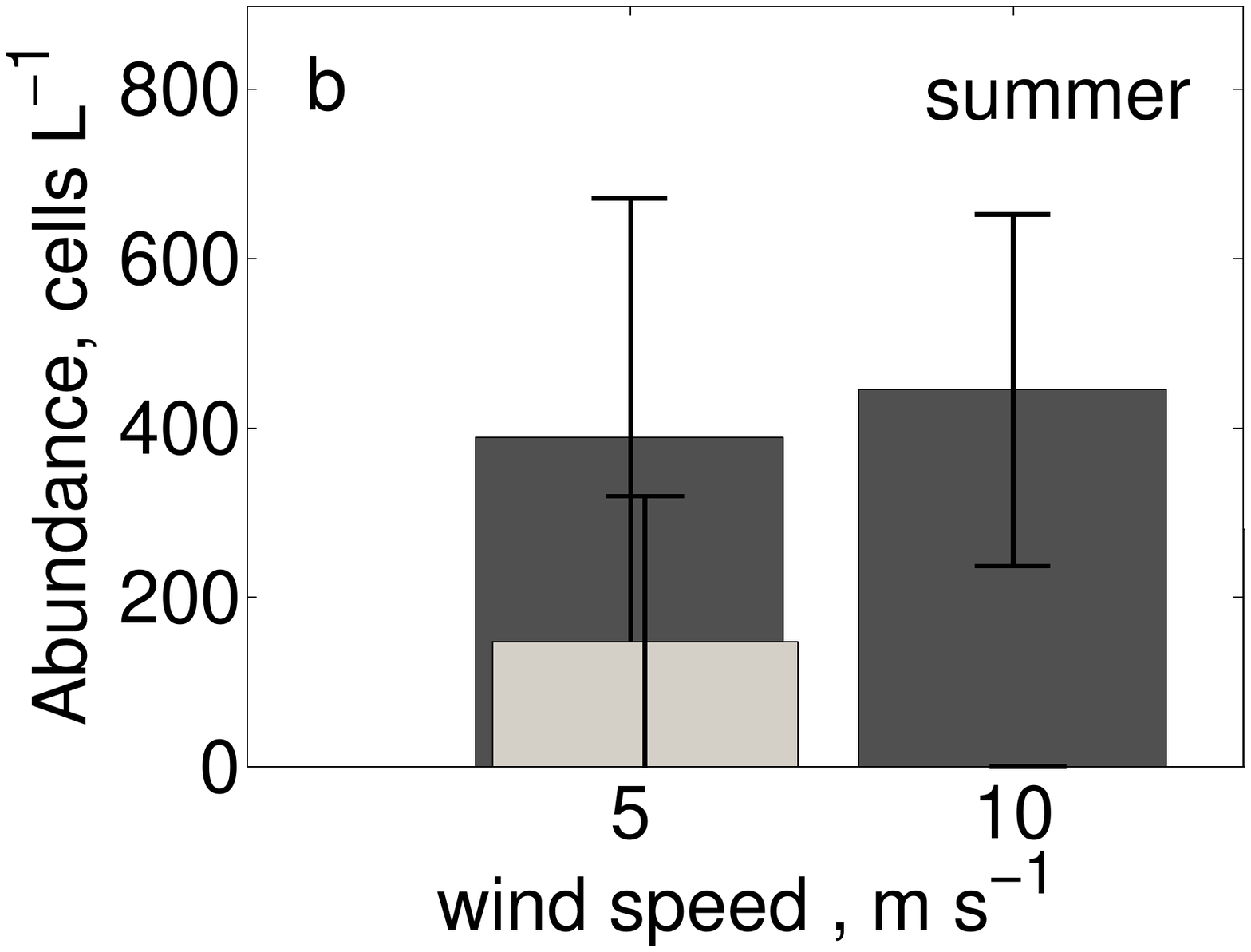}}
          \caption{Data sampling period is $16.10.2007-29.10.2009$. Mean and standard deviations of winter surface (light gray) and bottom (dark gray) abundances of \textit{P. sulcata} for winter (\textbf{a}) and summer (\textbf{b}) for two wind ranges: $(1)$ $[0, 5]$ m s$^{-1}$, $(2)$ $[5, 10]$ m s$^{-1}$. }\label{fig4}
        \end{figure}    
\subsection*{Long--term sampling of \textit{P. sulcata}}
The abundance of \textit{P. sulcata} in the water column typically remains below $7\times 10^3$ cells L$^{-1}$(the maximum abundance $~10^4$ cells L$^{-1}$ was observed in $2001$) (Fig. \ref{fig2} a). For the study period the mean winter densities of \textit{P. sulcata} ($1548$ cells L$^{-1}\pm 1340$ cells L$^{-1}$) by far exceeded summer densities ($171$ cells L$^{-1}\pm 485$ cells L$^{-1}$). The high winds were associated with the reduction of cell counts for both seasons (Fig.\ref{fig3}). 

\subsection*{Distributions of surface and bottom concentrations of \textit{P. sulcata}}

The $1$ m bottom cell counts generally exceeded the surface densities and ranged from $1500$ cells L$^{-1}$ to almost $3500$ cells L$^{-1}$ in winter and were typically below $1500$ cells L$^{-1}$ in the summer period. The surface cell densities did not exceed $2000$ cells L$^{-1}$ in winter and in summer were generally below $500$ cells L$^{-1}$ (Fig.\ref{fig4}).

According to analysis of diatom abundances at different depth calmer wind conditions were associated with higher cell counts at the surface. In contrast, an increase of wind led to the settling of part of algae to the sea bottom (Fig. \ref{fig4}): while only a slight decrease of cell counts occurred for winds $<5$ m s$^{-1}$ in winter (Fig. \ref{fig4}.a), in summer the surface concentrations decreased substantially (Fig. \ref{fig4}.b) which resulted in almost $100\%$ reduction of the surface cell counts (Fig. \ref{fig4}.b). 
 \subsection*{Simulation results}\label{sec8}
\begin{figure}
      \scalebox{0.4}{\includegraphics{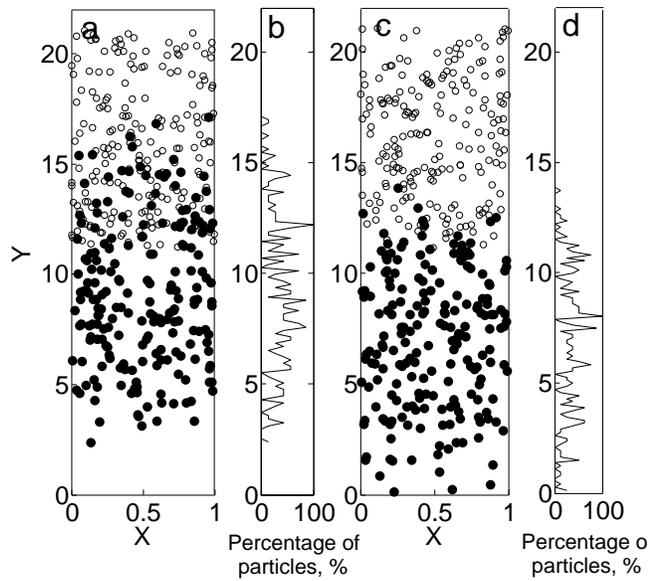}}
\caption{Simulation results from the model Eq. (\ref{equationMotion}). (\textbf{a}) Vertical profile of the distribution of particles at the initial time (empty circles) and at final time $t=20 \pi/\min(\omega_n)$ (filled circles) with parameters corresponding to winter values (\textbf{a}) and summer (\textbf{c}) values. (\textbf{b}) Percentage of particles found at a given distance  for winter (\textbf{b}) and summer (\textbf{d}). The X  in (\textbf{a}) and (\textbf{c}) is the width of domain in meters. }\label{fig5}
                  \end{figure}

The results of simulation for summer and winter parameters are shown in terms of two--dimensional vertical distribution of positions of $200$ particles (Fig. \ref{fig5} a and c) and the percentage of particles found at given depths (Fig. \ref{fig5} b and d). The initial positions of particles are chosen randomly in both cases. An average fall distance of the particles during time $t=369.01$ s in summer $\Delta Y_{sum}=9.14$ m was greater than in winter $\Delta Y_{win}=6.72$ m. From these estimates the average sinking speed was calculated: $v_{sum}=0.025
$ m s$^{-1}$ for summer and $v_{win}=0.018$ m s$^{-1}$ for winter. These values were substantially lower than the terminal velocities for the diatom--like particle of the same size: (\textit{i}) for winter $v_{win}=0.039$m s$^{-1}$ and (\textit{ii}) for summer $v_{sum}= 0.054$m s$^{-1}$ due to turbulent flow that slows in this case the vertical transport of particles. 
\section{Discussion}\label{sec9}
Experiments with artificial turbulence \citep{Ruiz2004} showed that intense mixing events impose environmental stress for phytoplankton cells. Therefore, not only turbulent mixing can homogenize the distribution of phytoplankton cells, but it can intensify the settling of cells in vitro \citep{Ruiz2004}. As in the natural condition, it was suggested that due to turbulent mixing more particles could be transported into trapping zone and further transferred to sediment \citep{Oliver2000}. Our results suggested that although an overall increase of cell concentrations in the water column was correlated with a higher wind intensity the algae vertical concentrations changed nonlinearly with the increase of wind. These changes were affected also by the seasonal water surface conditions as both field observations and model simulations here indicated. The phenomenon of a stress--avoiding behavior of diatoms \citep{Ruiz2004} remains to be observed in natural conditions. 

An increase of the bottom concentrations of diatoms at higher wind conditions could not be explained from the model described here. Most likely interaction of diatom cells with resuspended particulate matter could intensify diatoms sinking at higher winds.  
More complicated feedback between sediment material and water movement could change the character of the suspended sediment with attached chains of \textit{P. sulcata} diatoms. 

It was shown \citep{Chang2006} that storm and rough weather conditions in the southern North Sea and Wadden Sea could cause resuspension of heavy particles due to their break--up while sedimentation processes are typically observed more often during calm conditions. However, due to the lack of such observations for diatoms we do not discuss here the effects of break--up and clustering of particles on settling and resuspension processes \citep{Nichols1985}.  

Our theoretical estimates predicted higher sinking speed of diatom--like particles with increasing temperature due to lower sea water density and viscosity. Similar experimental studies of the sinking of diatoms \textit{Ditylum brightwelli} \citep{Bienfang1982} showed an evidence of the decrease of the sinking rate with increasing temperature. These experiments pointed to the existence of direct physiological response of diatoms to changes of temperature and salinity. In the model we did not account for the physiological responses. Detailed monitoring of vertical transport of \textit{P. sulcata} under different temperature, salinity and wind conditions could provide more insight into the existing mechanism of regulation of cell buoyancy under different physico--chemical and sea turbulence conditions. 
\section*{Conclusions}
Our analysis of long--term pelagic time series and weekly wind data for different seasonal conditions showed that stronger wind magnitude was associated with the reduction of diatom abundances in surface water. The decrease of abundances was less pronounced in winter. The reduction of cell counts was possibly linked to the changes in sinking of diatoms due to turbulent wave activity and changing buoyancy of algal cells. These observations were confirmed by separate analysis of the surface and bottom samplings of \textit{P. sulcata} that revealed more intense settling behavior of diatoms in summer than in winter at higher wind magnitudes. The numerical results showed similar behavior: the simulations performed with plankton--like particles inside two--dimensional turbulent field identified on average higher sinking speed for summer than for winter parameters. 
 \section*{Appendices}
The velocity field in Eq. (\ref{equationMotion}) is chosen according to the kinematic simulation model \citep{Fung2003} with a flow that models small--scale turbulence with a Kolmogorov $-5/3$ energy spectrum. This flow was used previously to investigate settling of heavy particles and mechanism of prolonged suspension of particles in a turbulent fluid \citep{Pasquero2003}.The two--dimensional incompressible turbulent velocity field is modeled as a sum of random incompressible Fourier modes:   
\begin{equation}
\mathbf{u}(\mathbf{x},t)=\sum_{n=1}^{N_k} [\mathbf{A}_n \cos(\mathbf{k}_n\cdot\mathbf{x} +\omega_n t)+\mathbf{B}_n \sin(\mathbf{k}_n\cdot\mathbf{x}+\omega_n t)],\label{velocityField}
\end{equation}  
$\mathbf{A_n}=A_n (\cos \phi_n,-\sin\phi_n ), \mathbf{B_n}=B_n (-\cos \phi_n,\sin \phi_n )$, $\mathbf{k_n}=k_n (\sin \phi_n,\cos\phi_n )$, and $N_k$ is the number of modes. The amplitudes $A_n$ and $B_n$ satisfy the condition: $A_n^2=B_n^2=E(k_n)\Delta k_n$. The condition of incompressibility $\mathbf{A_n}\cdot \mathbf{k_n}=\mathbf{B_n}\cdot \mathbf{k_n}=0$ of velocity field holds for all $n$. The angles $\phi_n$ are chosen randomly and uncorrelated with each other. Here Eulerian energy spectrum $E(k)$ is defined according to the expression: $E(k)=E_0 L(kL)^{-5/3}$, $L$ is the largest eddy scale of the turbulence. The wave numbers $k_n=k_1(L/\eta (N_k-1))^{n-1}$ are chosen according to geometric progression with $k_1=2\pi/L$ and $k_{N_k}=2\pi/\eta$. The frequencies $\omega_n$ are determined from the expressions $\omega_n=\lambda\sqrt{k_n^3 E(k_n)}$, where $\omega_n$ is proportional to the eddy turnover time of wave mode $n$. Here $\lambda$ is a dimensionless constant. 
 \section{Acknowledgements} \label{sec10}
We thank the crews of RV "Aade" and RV "Ellenbogen" for their efforts and dedicated work in the long--term monitoring program at Biologische Anstalt Helgoland. We are grateful to all colleagues who measured phytoplankton samples at HR. The authors acknowledge the German National Meteorological Service (Deutscher Wetterdienst) for the climate wind data. The work was supported by the German Federal Ministry for Education and Research (BMBF) [FK2 03F0609A,03F0603C].

\end{document}